\documentclass[times,twocolumn,final,authoryear]{elsarticle}

\usepackage{prletters}
\usepackage{framed,multirow}

\usepackage{amssymb}
\usepackage{latexsym}

\usepackage{url}
\usepackage{xcolor}

\usepackage{bm}
\usepackage{amsmath}
\usepackage{subfigure}
\usepackage{colortbl}
\usepackage{multirow}
\usepackage{booktabs}
\usepackage{multirow}

\definecolor{newcolor}{rgb}{.8,.349,.1}

\journal{Pattern Recognition Letters}

\begin{document}

\thispagestyle{empty}

\clearpage

\ifpreprint
  \setcounter{page}{1}
\else
  \setcounter{page}{1}
\fi

\begin{frontmatter}

\title{Bottom-up Broadcast Neural Network For Music Genre Classification}

\author[1]{Caifeng \snm{Liu}} 
\author[2]{Lin \snm{Feng}\corref{cor1}}
\cortext[cor1]{Corresponding author: 
  Tel.: +86-183-4222-9272;  }
\ead{fenglin@dlut.edu.cn}
\author[3]{Guochao \snm{Liu}}
\author[4]{Huibing \snm{Wang}}
\author[2]{Shenglan \snm{Liu}}

\address[1]{Faculty of Electronic Information and Electrical Engineering, Dalian University of Technology, Dalian 116024, China}
\address[2]{School of Innovation and Entrepreneurship, Dalian University of Technology, Dalian 116024, China}
\address[3]{Department of Functional R\&D, JD, Beijing 100176, china}
\address[4]{College of Information Science and Technology, Dalian Maritime University, Dalian 116024, China}


\begin{abstract}
Music genre recognition based on visual representation has been successfully explored over the last years. Recently, there has been increasing interest in attempting convolutional neural networks (CNNs) to achieve the task. However, most of existing methods employ the mature CNN structures proposed in image recognition without any modification, which results in the learning features that are not adequate for music genre classification. Faced with the challenge of this issue, we fully exploit the low-level information from spectrograms of audios and develop a novel CNN architecture in this paper. The proposed CNN architecture takes the long contextual information into considerations, which transfers more suitable information for the decision-making layer. Various experiments on several benchmark datasets, including GTZAN, Ballroom, and Extended Ballroom, have verified the excellent performances of the proposed neural network. Codes and model will be available at \emph{https://github.com/CaifengLiu/music-genre-classification}.

\end{abstract}


\end{frontmatter}


\section{Introduction}
\label{esc: Introduction}
With the rapid development of multimedia technology, a tremendous number of digital audios are uploaded on the Internet. Except the benefits brought by these audio tasks, the explosive growth of these audios causes fatal effects on various aspects. Therefore, managing these audios appropriately is a burdensome task crying out for reliable solutions. Researchers all over the world have devoted plenty of efforts to deal with various audios. Music information retrieval (MIR) is one of those studies, which provides a significative attempt to deal with music data. Due to the urgent need of many applications, such as music recommendation, music search, MIR has attracted attentions widely. Classification as a basic understanding of the music field has become an essential tool for MIR to analyze and process the music information. As a core issue of MIR, genre classification focuses on assigning a specific genre (classical, rock, jazz, etc.) to an unknown music clip. Expert annotation is notoriously expensive and intractable for large catalogues, Therefore, content-based genre recognition is highly valuable to bootstrap MIR system.

Even though researchers have proposed various algorithms from different perspectives, most of them rely on excellent hand-crafted features and constructing appropriate classifiers for music data. Spectrograms of music data have been proved to be one effective tool to describe audio signals. Similar to the images, spectrograms are also visual representations which can adequately maintain the information of time-frequency from music data. It builds a bridge between the algorithms for both image data and audio signals. Some mature algorithms for image processing can directly be adopted by related methods for audio signals. For the stage of feature extraction, the spectrograms are able to store most time-frequency information in their texture, which is of vital importance for the representation of various audios. Furthermore, there are many descriptors which can exploit the texture information to some extent, including Local Binary Patterns(LBP), Gabor Filters, Local Phase Quantization, etc.. After feature extraction, classification on the extracted features can directly determine the performances of music genre recognition. 
Some of those classification algorithms are common choices in this field, such as Support Vector Machine(SVM), Gaussian Mixture Models(GMM), and Music Classifier Systems(MCS) with different fusion strategies, etc. \citep{fu2011survey,wang2018beyond}.
Even though the feature representations and classification algorithms for music collections seem to be maturing, it's difficult for those traditional hand-crafted methods to design appropriate features for a specific task automatically. Therefore, it is far more beneficial to adopt a data-driven method rather than designing those hand-crafted ones.

Over the last decade, we have witnessed a surge of Convolutional Neural Network (CNN)  architectures which have achieved satisfying performances in many fields like image recognition \citep{wang2017effective,wu2019cycle} and natural language processing. Meanwhile, Music genre classification has also been inspired by the remarkable successes of CNNs. It is well known that CNNs can extract enough information from images due to their hierarchical structures \cite{wu2018deep}. Low-level features, such as underlying texture, etc.,  are constructed into high-level semantic information through all layers of CNNs. Similar to images, music also consists of hierarchical structures, which inspires us to develop an appropriate CNN model to deal with the problem of music classification. For instance, pitch and loudness combine over time to form chords, melodies, and rhythms. And all these elements above can form the whole music layer by layer. Furthermore, it has been verified that CNNs are very sensitive to the textural information \citep{hafemann2014forest} of images. The special ability of CNNs can help the task of music classification to exploit abundant information from spectrogram which contains rich texture information from music signals.

Up to now, most of the CNN-based music classification models are constructed by directly utilizing those mature architectures to deal with this problem. For example, Choi \emph{et al.} \citep{choi2017transfer} introduced a musical transfer learning system, in which a CNN model was trained on a large music dataset \citep{bertin2011million} as a feature extractor and an SVM classifier was stacked on it and Jakubik \emph{et al.} \citep{jakubik2017evaluation} introduced two Recurrent Neural Network(RNN) architectures from image domain with a different mechanism of gating: Long-Short Term Memory(LSTM) and Gated Recurrent Unit(GRU). They reported the accuracies of 89\% and 92\% respectively on GTZAN, which showed their potential for music content analysis. 

Even though these architectures have achieved excellent performances in the music domain, the results have not been nearly as convincing as they have been in the visual realm. Most CNN-based models directly fed the visual spectrogram representations into these CNNs without any modification. Because traditional CNNs are constructed just for image processing tasks, it cannot achieve good performances without awareness of the difference between spectrogram and images. Stimulated by the problems above, two main motivations of our paper are listed as follows:

\begin{itemize}
    \item Even though the genres are positioned in different levels or time-scales in each hierarchy previous deep learning based works predict the genre from the same scale of time and level frequency, which is similar with the task of image classification. However, sound events are accumulated by frequency over time domain which causes the individual genre has different performance sensitivity to different time scales and levels of features. Therefore, it is necessary to design a specific CNN structure which can comprehensively handle multi-scale of audio features. 
    \item Previous network structures of music genre classification mainly focus on abstracting high-level semantic features layer by layer. It leads to a massive loss of lower-level features which include a large amount of critical information for making the decision. However, the low-level features tend to be more contributed for improving the genre classification performance \citep{choi2017transfer}. Therefore, how to construct an appropriate CNN structure to maximally abstract high-level information and preserve the lower-level features simultaneously, which is just for the task of music classification is of vital importance but challenging.
\end{itemize}

\begin{figure}[htbp]
    \label{Fig:Comparison of different blocks}
    \centering
    \subfigure[Dense Connectivity]{
    \begin{minipage}[t]{0.4\linewidth}
    \label{Fig:Dense Block}
    \centering
    \includegraphics[width=1.6in]{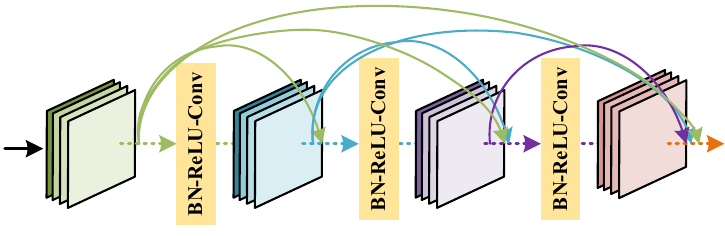}
    \end{minipage}
    }
    \hfill
    \subfigure[Inception Block]{
    \begin{minipage}[t]{0.5\linewidth}
    \label{Fig:Inception Block}
    \centering
    \includegraphics[width=1.85in]{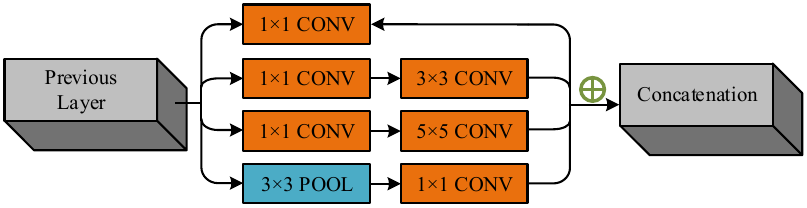}
    \end{minipage}
    }
     \subfigure[Broadcast Module]{
     \label{Fig:Broadcast Module}
     \centering
     \includegraphics[width=3.6in]{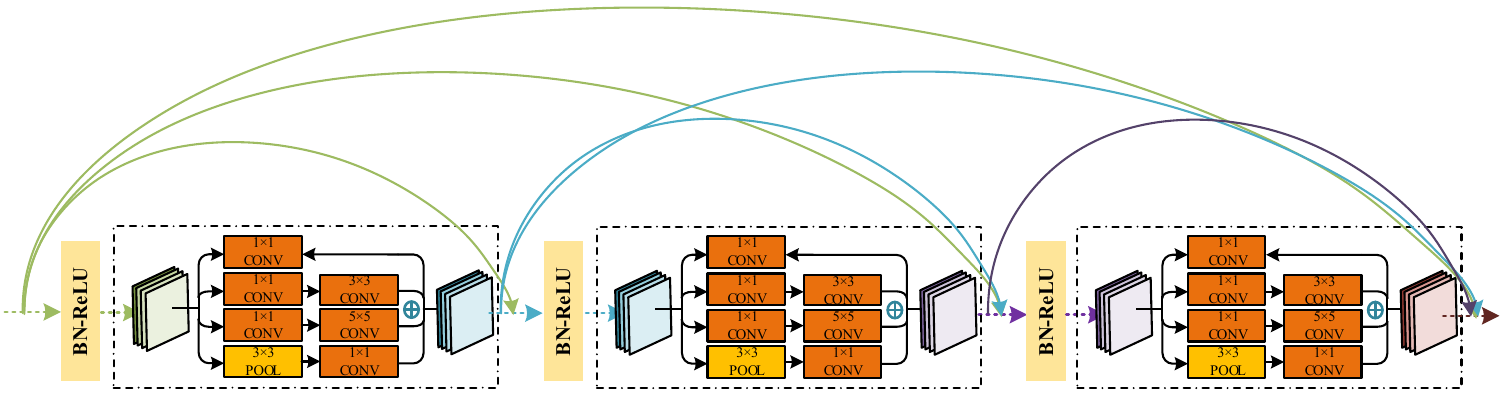}
    }

    \caption{Comparison of prior network structures (a, b) and proposed module Module(c). (a) Dense Block. (b) Inception Block. (c) Proposed Broadcast Module}
\end{figure}

Based on the analysis above, the direct way to deal with these problems above is to exploit an appropriate CNN model which can make full use of both the high-level semantic information and the low-level features from various music. In this paper, we surveyed the problems in the field of music classification and proposed a novel architecture named Bottom-up Broadcast Neural Network (BBNN) which adopts a relatively wide and shallow structure. The main idea of the BBNN architecture is to develop effective block and connection manner between different blocks to fully exploit and preserve the low-level information to the higher layers. So that, low-level information of spectrogram is able to participate in the decision-making layer throughout the network, which is very important for the task of music classification. Therefore, BBNN is equipped with a novel Broadcast Module (BM) which consists of Inception blocks and connects them by dense connectivity. We have shown the architecture of BM in Figure~\ref{Fig:Broadcast Module}. Because the Inception block can perceive the feature maps with different scales, it is able to extract information embedded in the time-frequency of the audio signal from different scales simultaneously. Moreover, BBNN densely connects those basic blocks and transforms the low-level information to the decision-making layers, which ensures that the low-level information can be maintained as much as possible. Most Deep CNN (DCNN) models have to adopt various data-augmentation pre-processings \citep{salamon2017deep} to enlarge the size of the training dataset. Compared with those traditional DCNN models, our proposed BBNN has few parameters to be learned. Therefore, a smaller dataset without any data-augmentation techniques, such as Ballroom, is enough for the training stage of BBNN.  

\section{Proposed Design And Approach}
\label{Sec: Design and Approach}

\subsection{Broadcast Module}
\label{Sec: Broadcast Module}

The widely accepted consensus is that individual genre has different performance sensitivity to various frequency bands and time intervals. Inspired by \citep{szegedy2015going}, we combine convolutions with different kernel size to form an Inception block (Figure~\ref{Fig:Inception Block}).
The Inception blocks are stacked on top of each other as basic extraction unit to sufficiently learn features from multiple reception fields. It can decrease the network susceptibility to frequency-shifts in a spectrogram. 
To further strengthen feature propagation in the BM, we utilize dense connection paths to connect all Inception blocks in BM, which can bottom-up transmit the extracted feature maps to all subsequent blocks in BM.
One of the main beneficial aspects of BM architecture is that it maximally transmit and preserve all extracted feature maps to higher-layers so that the decision layers make a prediction based on all feature-maps in the network. 
Another practically useful aspect of BM design is that it aligns with the intuition that audio information should be perceived at various time-frequency scales simultaneously.

As shown in Figure~\ref{fig:BBNNarchitechture},
BM consists of $L$ identical Inception blocks connected each other by dense connectivity, which allows each block to receive inputs directly from all its previous blocks. We denote ${{\bf{X}}_{SL}}$ outputted by shallow layers as the input of the BM, $L$ as the number of Inception blocks. In the music genre recognition task, we fix the $L=3$. Thus, the input of $l$-$th$ block, $l = 1,...,L$, can be represented as:
\begin{equation}
    {{\bf{X}}_l} = {f_I}\left( {\left[ {{{\bf{X}}_{SL}},{{\bf{X}}_1},...,{{\bf{X}}_{l - 1}}} \right]} \right),
\end{equation}   
where ${\left[ {{{\bf{X}}_{SL}},{{\bf{X}}_1},...,{{\bf{X}}_{l - 1}}} \right]}$ refers to the concatenation of the feature maps produced by blocks $0,...,l - 1$ and ${f_I}$ is a composite function of all operations in the Inception block. In each Inception block, the filter sizes of convolutions are mainly adopted $1 \times 1$, $3 \times 3$, $5 \times 5$ with stride 2 and then, $1 \times 1$ convolutions are utilized to compute reductions before them. 
Before each convolution, the BN and rectified linear activation operations are implemented.
An Inception block consists of layers of above types stacked upon each other, with occasional max-pooling layers restricted to the size $3 \times 3$ with stride 2 to halve the resolution of the grid. 
The use of the Inception block is based on \citep{szegedy2015going}, although our implementation differs in that we employ an extra BN layer before each convolution.
This makes the network generalization ability significantly enhanced even trained on a small-scale dataset.
As shown in the Table~\ref{Tab:architecture parameters}, the growth rate $k$ of the BM is 128. Each block has ${k_0} + k \times \left( {l - 1} \right)$ input feature maps, where ${k_0}$ is the number of channels in the input ${{\bf{X}}_{SL}}$. The exact BM configurations used in the experiment are shown in Table~\ref{Tab:architecture parameters}. For illustration purpose, we divide the structure of Inception block into a top and bottom parts and list them respectively.

\subsection{Network Structure}
\label{sec: Network Architecture}

\begin{figure}[!t]
\begin{center}
\begin{tabular}{c}
\includegraphics[height=8cm]{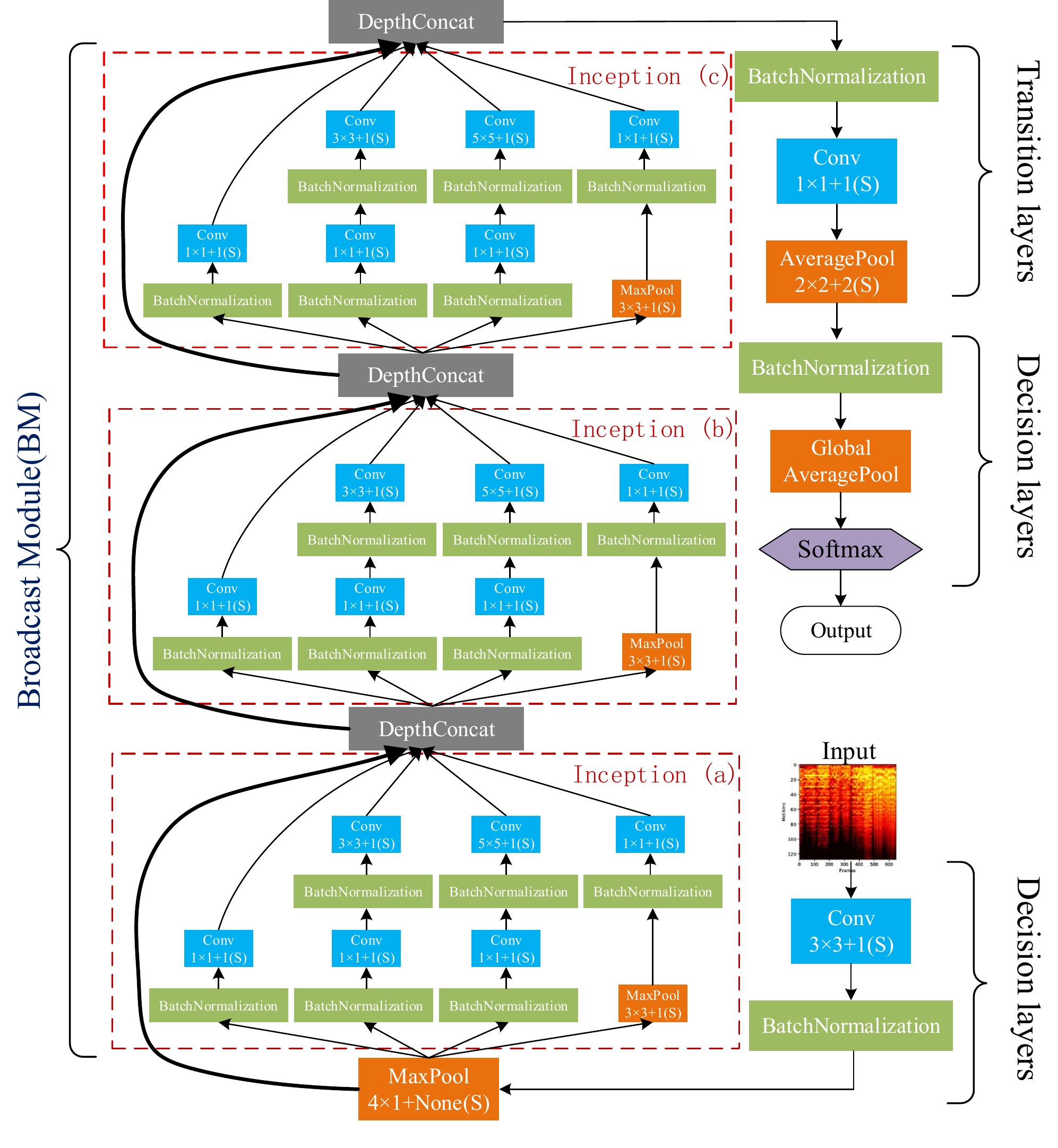}
\end{tabular}
\end{center}
\caption
{ \label{fig:BBNNarchitechture}
The corresponding network architecture of BBNN }
\end{figure}

As shown in Figure~\ref{fig:BBNNarchitechture}, The BBNN comprises 9 layers when counting only layers with parameters (or 12 if we also count pooling layers). Each of layers implements a non-linear transformation such as Convolution (Conv), Softmax, Batch Normalization (BN) \citep{ioffe2015batch}, and Pooling operation. Inspired by \citep{ioffe2015batch}, We execute the BN transform immediately after each convolution operation and then use rectified linear activation (ReLU). A main beneficial aspect of BN is that it regularizes our model and reduces the need for Dropout.

All layers of the proposed network can be summarized in
four parts to play different roles as follows: shallow feature extraction layers, BM, transition layers, and decision layers. 
The whole model aims to learn the all parameters $\Theta $ of a composite function $F\left( {\left.  \cdot  \right|\Theta } \right),$ which maps input ${{\bf{X}}_0}$ to the output (genre) $p$:
\begin{equation}
  p = F\left( {\left. {{{\bf{X}}_{\bf{0}}}} \right|\Theta } \right) = {f_{DL}}\left( {\left. {{f_{TL}}\left( {\left. {{f_{BM}}\left( {\left. {{f_{SL}}\left( {\left. {{{\bf{X}}_{\bf{0}}}} \right|{\theta _{SL}}} \right)} \right|{\theta _{BM}}} \right)} \right|{\theta _{TL}}} \right)} \right|{\theta _{DL}}} \right),
\end{equation} 
where the index of $f\left(  \cdot  \right)$ represents a composite function of corresponding part of network.
Specifically, the shallow layers (the ones close to the input) include a $3 \times 3$ convolution, a BN, and a $4 \times 1$ max pooling with 1 stride.
A relatively small receptive field is used to extract the local frequency information in a short time span. After activating the local features with a BN and ReLU functions, we further add a max-pooling operation. It is considered that human is more concerned about the salient tempo in a short time when they recognizing the music genre.
The max-pooling layer can filter out the dominant frequency in the short time interval of the mel-spectrogram. Furthermore, it makes the model possess some capacity of translation invariance.    
According to the above operations, 
the extracted local information is transmitted into each layer of BM and fused to gather evidence in support of contextual "time-frequency signatures" that are indicative of recognizing different musical genres. The structural details about BM were given in Section~\ref{Sec: Broadcast Module}.

Down-sampling layer is an essential part of convolutional networks. After extracting hierarchical features with BM, we further conduct several down-sampling layers to reduce the size of feature-maps and the number of channels which are significantly increased by the concatenation operations used in BM. These layers between BM and decision layers are referred to as transition layers, which do a BN, ReLU activation, $1 \times 1$convolution and $2 \times 2$ average-pooling with stride 2 operations. 

At the final decision stage of BBNN, instead of adding fully connected layers stacked on the feature maps \citep{jakubik2017evaluation}, we utilize global average pooling  \citep{lin2013network} layer to take the average of each feature map. It is easier to interpret the correspondence relations between feature maps and genres and less prone to overfitting than traditional fully connected layers.   
Then, the resulting vector of the global average pooling layer is fed into a softmax log-loss function which can produce a distribution over the genre labels (blues, classic, etc.).

The BBNN is designed with full consideration of computational efficiency and practicality. Here, the configurations of BBNN is described in Table~\ref{Tab:architecture parameters} for demonstrating the specific architectural parameters, where the size of mel-spectrogram ${{\bf{X}}_0}$ is $647 \times 128$ (30s music duration). In all convolution layers, we pad zeros to each side of the input to keep size fixed.
As seen from Tabel~\ref{Tab:architecture parameters}, the trained model has a tiny size, only 2.4M, which can be applied in individual devices including even those with limited computational resources.   

\definecolor{mygray}{gray}{.9}
\begin{table}[!htbp]
    \centering
    \tiny
    \caption{The configuration of BBNN. Note that each convolution layer shown in the table corresponds the sequence BN-ReLU-Conv.}
    \label{Tab:architecture parameters}
    \begin{tabular}{|c|c|c|c|c|}
        \hline
        \rowcolor{mygray}
        Type & Layers & Output Size & Filter Size/Stride (Number) & Params \\
        \hline
        \multirow{2}*{SL}&Convolution& $647 \times 128 \times 32$ & $3 \times 3/1(32)$ & 320\\
        \cline{2-5}
        & Max Pool & $161 \times 128 \times 32$ & $4 \times 1/None$ & \\
        \hline
        \multirow{9}*{BM}&Inception (a), top&- & $\left[ {1 \times 1/1(32){\rm{conv}}} \right]*3$, $\left[ {3 \times 3/1{\rm{max pool}}} \right]*1$ & 3,168\\
        \cline{2-5}
        &\multirow{2}*{Inception (a), bottom}
        &\multirow{2}*{$161 \times 128 \times 160$}
        & $\left[ {3 \times 3/1(32){\rm{conv}}} \right]*1$, $\left[ {5 \times 5/1(32){\rm{ conv}}} \right]*1$
        &\multirow{2}*{35,936} \\
        &\multirow{2}*{}&\multirow{2}*{}&$\left[ {1 \times 1/1(32){\rm{conv}}} \right]*1$ &\\
        \cline{2-5}
        &Inception (b), top& - &$\left[ {1 \times 1/1(32){\rm{conv}}} \right]*3$, $\left[ {3 \times 3/1{\rm{max pool}}} \right]*1$ & 15,456\\
        \cline{2-5}
        &\multirow{2}*{Inception (b), bottom}
        &\multirow{2}*{$161 \times 128 \times 288$}
        & $\left[ {3 \times 3/1(32){\rm{conv}}} \right]*1$, $\left[ {5 \times 5/1(32){\rm{ conv}}} \right]*1$
        &\multirow{2}*{40,032} \\
        &\multirow{2}*{}&\multirow{2}*{}&$\left[ {1 \times 1/1(32){\rm{conv}}} \right]*1$ &\\
        \cline{2-5}
        &Inception (c), top& - &$\left[ {1 \times 1/1(32){\rm{conv}}} \right]*3$, $\left[ {3 \times 3/1{\rm{max pool}}} \right]*1$ & 27,744\\
        \cline{2-5}
        &\multirow{2}*{Inception (c), bottom}
        &\multirow{2}*{$161 \times 128 \times 416$}
        & $\left[ {3 \times 3/1(32){\rm{conv}}} \right]*1$, $\left[ {5 \times 5/1(32){\rm{ conv}}} \right]*1$
        &\multirow{2}*{44,128} \\
        &\multirow{2}*{}&\multirow{2}*{}&$\left[ {1 \times 1/1(32){\rm{conv}}} \right]*1$ &\\
        \cline{1-5}
        \multirow{2}*{TL}& Convolution & $161 \times 128 \times 32$ & $1 \times 1/1(32)$ & 13,344 \\ 
        \cline{2-5}
        & Max Pool & $80 \times 64 \times 32$ & $2 \times 2/2$ &  \\
        \hline
        \multirow{2}*{DL}& Global Average Pool & $1 \times 1 \times 32$ & - &  \\
        \cline{2-5}
        & Softmax & $1 \times 1 \times 10$ &- & 330 \\
        \hline
        \hline
        \multicolumn{4}{|c|}{\textbf{Total Params}} &\textbf{180,458}\\
        \hline

        

     \end{tabular}

\end{table}

\section{Experiments}
\label{sec: Experiment}
\subsection{Datasets}
\label{sec:Datasets}
\textbf{GTZAN.} The dataset has been widely used in many studies with the aim of music genre classification. It was collected and proposed by Tzanetakis in \citep{tzanetakis2002musical}. The genre labels and numbers of corresponding genres are given in Table~\ref{Tab: Dataset description}.

\textbf{Ballroom}. The dataset
\citep{cano2006ismir} consists of clear and constant rhythmic patterns, which makes it suitable for recognition tasks.
The specific genres and the corresponding number of each genre are listed in Table~\ref{Tab: Dataset description}.

\textbf{Extended Ballroom}. The dataset \citep{marchand2016extended} was proposed in 2016 by Marchand, which extended the original Ballroom dataset. Comparing to the original one, the extended version contains six times more tracks of better audio quality. We show in Table~\ref{Tab: Dataset description} the genre class distribution of the dataset. The imbalance of this dataset poses vast challenges for genre classification.

\begin{table}[!thbp]
\centering
\tiny
\caption{Datasets Description}
\label{Tab: Dataset description}
\begin{tabular}{|l c||l c||l c|}
     \hline
     \rowcolor{mygray}
     \multicolumn{2}{|l||}{GTZAN}
     &\multicolumn{2}{l||}{Ballroom} &\multicolumn{2}{l|}{Extended Ballroom}\\
     \hline
     Genre & Track & Genre & Track & Genre & Track \\
     \hline
     Classic & 100 &Cha Cha &111 & Cha Cha & 455 \\
     Jazz & 100 &Jive & 60 & Jive & 350 \\
     Blues & 100 &Quickstep &82 & Quickstep & 497 \\
     Metal & 100 &Rumba &98 & Rumba & 470 \\
     Pop & 100 &Samba &86 & Samba & 468 \\
     Rock & 100 &Tango &86 & Tango & 464 \\
     Country & 100 & Viennese Waltz &65 & Viennese Waltz & 252 \\
     Disco & 100 &Slow Waltz &110  & Waltz & 529 \\
     Hiphop & 100 & & & Foxtrot & 507 \\
     Reggae & 100 & & & Pasodoble & 53 \\
      & & & & Salsa & 47 \\
      & & & & Slow Walz & 65 \\
      & & & & Wcswing & 23 \\
     \hline
     \hline
     \textbf{Total} & \textbf{1000} & \textbf{Total} & \textbf{698} & \textbf{Total} & \textbf{4180}  \\
     \hline
     
\end{tabular}
\end{table}
\subsection{Preprocessing}
\label{perprocessing}
In this work, mel-spectrogram is utilized as input to the proposed network.
Specifically, we use Librosa \citep{mcfee2015librosa} to extract mel-spectrograms with 128 Mel-filters (bands) covering the audible frequency range (0-22050 Hz), setting a frame length of 2048 and a hop size of 1024. We can get mel-spectrogram of size $647 \times 128$.

\subsection{Training and other details}
\begin{figure}[!thbp]
    \centering
    \begin{tabular}{c}
        \includegraphics[height=2.5cm]{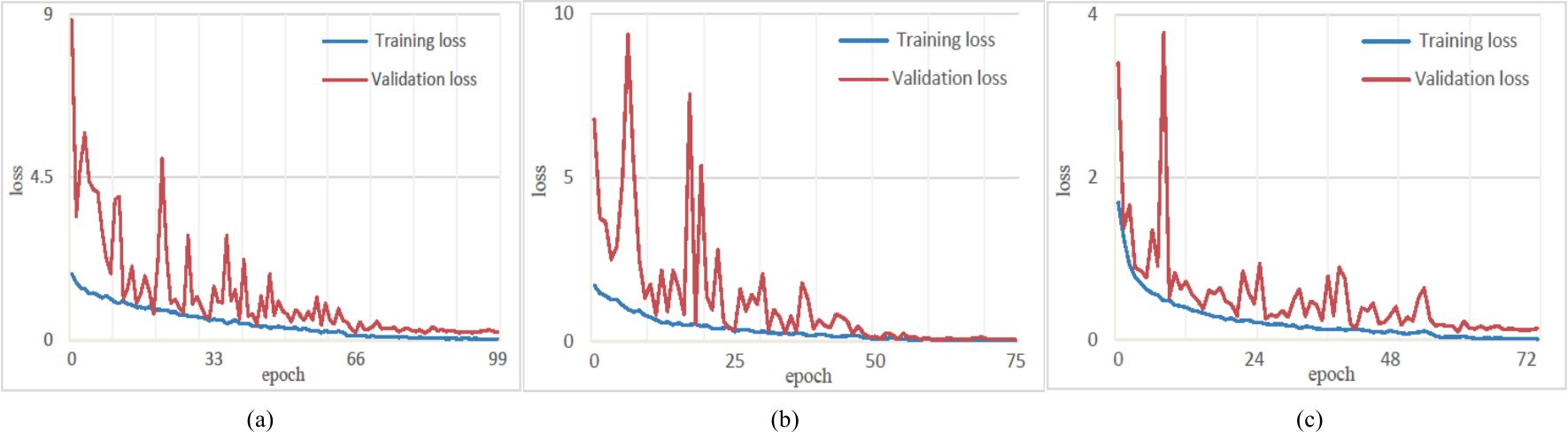}
    \end{tabular}
    \caption{Loss curves of training and validation on the different datasets: (a) GTZAN, (b) Ballroom, (c) Extended Ballroom}
    \label{fig:Loss curves }
\end{figure}

All files for each dataset are transformed to mel-spectrograms by the preprocessing program presented in Section~\ref{perprocessing}. The mel-spectrogram with size $647 \times 128$ input to the BBNN.
All the models are trained to minimize categorical cross-entropy between the predictions and truthful genre labels utilizing ADAM \citep{kingma2014adam} optimizer. All the three datasets use batch size 8 for 100 epochs. We set initial learning rate is 0.01 and automatically decrease it by a factor of 0.5 when the loss has stopped improving after 3 epochs.
In addition, we set up an early stop mechanism, that is, training stops when a monitored quantity has stopped improving even if the epoch does not reach 100. Figure~\ref{fig:Loss curves } shows the training and validation loss curves of BBNN network on GTZAN, Ballroom, and Extended Ballroom datasets. BBNN converges to a low loss whether training set or verification set. We further analyze the BBNN's effect on the test sets of each dataset in more detail below. \

\textbf{Metric.}
Following previous works (e.g. \citep{choi2017transfer}), we perform a 10-fold cross validation to evaluate the classification accuracy across all experiments. The training, testing, and validating sets are randomly partitioned following proportion 8/1/1. The total classification accuracy is calculated as the average of 10-folds cross-validations.

\textbf{Experiment Platform.}
Our code is written by Python, based on the Keras \cite{chollet2015keras} and the publicly available toolbox of preprocessing Librosa \cite{mcfee2015librosa}. All of our experiments are running on NVIDIA TITAN Xp GPU with 12 GB memory.

\subsection{Classification results on GTZAN}
\label{sec:Classification results on GTZAN}
In Table~\ref{tab:GTZAN}, we compare BBNN with some recent excellent models including 6 different deep learning models and 1 traditional method based on a hand-crafted feature descriptor. Audeep  \citep{freitag2017audeep} was based on a recurrent sequence to sequence autoencoder, which took full consideration of temporal dynamics of audio data and yielded an accuracy of 85.4\%. The transform learning framework \citep{choi2017transfer} obtained an accuracy of 89.8\%, which is transplanted to the music genre classification task from the visual domain. 
Hybrid model, CVAF and MFMCNN relied on different strategies of feature fusion to improve classification accuracy and generated the accuracies of 88.3\%, 90.9\% and 91.0\% respectively. Multi-DNN generated a slightly lower accuracy than BBNN by using a framework cascaded multiple DNN networks which consumes more resources and strongly depends on an additional database to train the model. 

\begin{table}[htbp]
    \centering
    \tiny
    \caption{Classification accuracy (\%) on GTZAN dataset is compared across recently proposed methods (the best result is marked in bold). the results of all methods have reported in the original papers or related literatures}
    \label{tab:GTZAN}
    \begin{tabular}{ccc}
        \hline
        Methods & Preprocessing & Accuracy(\%) \\
        \hline
            AuDeep \citep{freitag2017audeep} & mel-spectrogram & 85.4 \\

            NNet2 \citep{zhang2016improved} & STFT & 87.4 \\
            Hybrid model \citep{karunakaran2018scalable} & MFCC, SSD, etc. & 88.3 \\
            Transform learning \citep{choi2017transfer} & MFCC & 89.8 \\
            CVAF \citep{nanni2017combining} & mel-spectrogram, SSD, etc. & 90.9 \\
            MFMCNN \citep{senac2017music} & STFT, ZCR, etc. & 91.0 \\
            Multi-DNN \citep{dai2015multilingual} & MFCC & 93.4 \\
            \hline
            \textbf{Ours} & mel-spectrogram & \textbf{93.9} \\
        \hline
    \end{tabular}
\end{table}

Figure~\ref{fig: confusions}(left) shows the confusion matrix of 10-folds results predicted by BBNN on GTZAN. The rows and columns of the matrix represent ground-truths and their predicted labels.
The diagonal numbers of matrix respectively represent correct prediction per genre and the off-diagonal entries are confusions between different genres.  
Confusion matrix can give the individual discrimination relation between the ground-truth and predicted genre label, which provides a better view of the general classification performance of the BBNN model. Table~\ref{tab:precisions and recall on GTZAN} lists the precision, recall rate and F-score of each genre corresponding the confusion matrix.
From the Figure~\ref{fig: confusions}(left) we can see that the proposed model distinguishes most of the genres very well, but it is highly confused to distinguish Rock from Country and Metal. One explanation is that they might share more similar frequency information which makes it more difficult to classify them in nature. Nonetheless, expert advice probably is required to improve the classification accuracy on the Pop and Rock genres. It is further discussed in Section~\ref{sec:conclusions and future work}. Overall, as shown by the Table~\ref{tab:precisions and recall on GTZAN}, it appears to that most genres have been correctly classified and recall rate and precision of genre Jazz even reach 99\%.  



\begin{table}[htbp]
\centering
\tiny
\caption{Precision (\%) and recall rate (\%) of each genre obtained on the GTZAN dataset}
\label{tab:precisions and recall on GTZAN}
\begin{tabular}{cccc}
    \hline
    Genre & Precision (\%) & Recall Rate (\%) & F-score (\%) \\
    \hline
    Blues & 90.8 & 97.0 &93.8 \\
    Classical & 98.9 & 97.9 & 98.4\\
    Country & 89.8 & 97.7 & 93.6\\
    Disco & 98.2 & 91.8 & 94.9\\
    Hip-hop & 93.0 & 94.9 & 93.9\\
    Jazz & 99.0 & 99.0 & 99.0\\
    Metal & 86.1 &98.7 &92.0 \\
    Pop & 94.3 & 94.3  & 94.3\\
    Reggae & 94.2 & 88.1 & 91.1\\
    Rock & 93.3 & 80.7 & 86.6 \\
    \hline
    \textbf{Average} & \textbf{93.7} & \textbf{94.0} & \textbf{93.7} \\
    \hline
\end{tabular}
\end{table}

\subsection{Classification results on Ballroom}
Table~\ref{tab:Ballroom} shows the classification accuracy percentage results obtained on Ballroom dataset by BBNN framework and 5 novel methods including 3 deep learning frameworks (MMCNN, MCLNN, Pons \emph{et al.} \citep{pons2017designing}) and two traditional methods based on different hand-crafted feature representations. 
The work \citep{pons2017designing} presented by Marchand \emph{et al.} is based on Modulation Scale Spectrum presentation of audio (called MSS) and a modified KNN classifier is used to perform the classification achieving an accuracy of 87.6\%. Based on MSS, Marchand \emph{et al.} then proposed a Modulation Scale Spectrum with Auditory Statistic representation (SOTA) and used an SVM as the classifier, which boosted the recognition accuracy by about 3\%. The MMCNN architecture has two layers (CNN $+$ feed-forward) and uses two different filter shapes in the CNN layers (1-by-60 and 32-by-1). It produces more parameters (196,816) than BBNN to build the classification model and generates relatively lower accuracy. To preserve the inter-frames relation of a temporal signal, Medhat \emph{et al.} designed a masked conditional neural network (MCLNN) which obtained an accuracy of 90.4 \%. Pons \emph{et al.} paid attention to temporal features of audio to use wider kernels of convolution layers that span over the long time duration. They used fewer parameters (92,808) than BBNN for modelling the network but generated a relatively lower accuracy than SOTA and BBNN. For this dataset, our proposed BBNN network (accuracy of 97.1\%) outperforms all the compared models. 
\begin{table}[htbp]
    \centering
    \tiny
    \caption{Classification accuracy (\%) on Ballroom dataset is compared across recently proposed methods (the best result is marked in bold). the results of all methods have reported in the original papers or related literatures. Note the exception that SOTA is reported in recall rate.}
    \label{tab:Ballroom}
    \begin{tabular}{ccc}
        \hline
        Methods & Preprocessing & Accuracy(\%) \\
        \hline
        MMCNN \citep{pons2016experimenting} & mel-spectrogram & 87.6 \\
        MCLNN \citep{medhat2017automatic} & mel-spectrogram & 90.4 \\   
        Pons \emph{et al.} \citep{pons2017designing} & mel-spectrogram & 92.1 \\       
        Marchand \emph{et al.} \citep{marchand2014modulation} & MSS & 93.1 \\
        SOTA \citep{marchand2016scale} & MASSS & 96.0 \\
        \hline
        \textbf{Ours} & mel- spectrogram & \textbf{96.7} \\
        \hline
    \end{tabular}
\end{table}

Figure~\ref{fig: confusions}(middle) illustrates more detailed information about the BBNN performance in the form of a confusion matrix. The corresponding precision and recall rate for each genre is further listed in the Table~\ref{tab:precisions and recall on Ballroom}.
Through the confusion matrix, it is clear exhibited that the strong ability of BBNN to recognize the most genres, for example, Cha Cha and Quickstep. There are relatively easy confusions between the Rumba and Slow Waltz genres. This is due to the reason that the genre boundaries of Rumba and Slow Waltz are not clear cut as Cha Cha or Quickstep.
Such confusion leads to their relatively lower precisions and recall rates than other genres as shown in the Table~\ref{tab:precisions and recall on Ballroom}. It is noticeable that the BBNN model can achieve good performances in both precision and recall rate for almost genres. Average recall rate is 0.9 \% higher than SOTA. 


\begin{table}[htbp]
\centering
\tiny
\caption{Precision (\%) and recall rate (\%) of each genre obtained on the Ballroom dataset}
\label{tab:precisions and recall on Ballroom}
\begin{tabular}{cccc}
    \hline
    Genre & Precision (\%) & Recall Rate (\%) & F-score (\%)\\
    \hline
    Cha Cha & 100 & 96.2 & 98.0 \\
    Jive & 97.9 & 94.1  & 96.0\\
    Quickstep & 96.4 & 100 &98.1 \\
    Rumba & 97.0 & 94.2 & 95.6 \\
    Samba & 95.0 & 97.4 & 96.2 \\
    Tango & 98.8 & 98.8 & 98.8 \\
    Viennese Waltz & 98.5 & 94.3 & 96.4 \\
    Slow Waltz & 94.3 & 100 & 97.0 \\
    \hline
    \textbf{Average} & \textbf{97.2} & \textbf{96.9} & \textbf{97.0} \\
    \hline
\end{tabular}
\end{table}

\subsection{Classification results on Extended Ballroom}
\label{sec:Classification results on Extended Ballroom}
Table~\ref{tab:extended ballroom} reports the comparison of BBNN with new state-of-the-art music genre classification methods in term of accuracy. 
BBNN achieves an accuracy of 97.2\% on this dataset, which surpasses the methods including CNN-based architectures \citep{choi2017transfer,jeong2017dlr,pons2018randomly} and hand-crafted feature approach \citep{marchand2016scale}. 
Different representations of audio signals including MFCC, mel-spectrogram and MASSS represent preprocessing programs of audio signals utilized in the corresponding methods. The first work \citep{choi2017transfer} proposed by Choi \emph{et al.}, achieved an accuracy of 86.7\% using a transfer learning framework. Specifically, a Vggnet was designed and trained for a source dataset including 244,224 music clips, then, the trained model was adopted to Extended Ballroom dataset as a feature extractor. DLR \citep{jeong2017dlr} also is a transform learning framework as similar as the abovementioned work \citep{choi2017transfer}, but the difference is that DLR aims to learn a rhythmic representation for a source task which can be used as an input for musical genre recognition task. Compared with these transform learning frameworks, BBNN generates a higher accuracy and does not require pre-training on the larger dataset. In RWCNN, randomly weighted CNN architecture was utilized to extract features for a classifier (e.g. SVM and ELM). Its accuracy is relatively lower than BBNN. The comparative experiment results strongly validate the effectiveness of BBNN.

\begin{table}[htbp]
    \centering
    \tiny
    \caption{Classification accuracy (\%) on Extended Ballroom dataset is compared across recently proposed methods (the best result is marked in bold). Note the exception that SOTA is reported in recall rate. the results of all methods have reported in the original papers or related literatures}
    \label{tab:extended ballroom}
    \begin{tabular}{ccc}
        \hline
        Methods & Preprocessing & Accuracy(\%) \\
        \hline
        Transform learning \citep{choi2017transfer} & MFCC & 86.7  \\
        RWCNN \citep{pons2018randomly} & MFCC & 89.8 \\
        DLR \citep{jeong2017dlr} & mel-spectrogram & 93.7 \\
        SOTA \citep{marchand2016scale} & MASSS  & 94.9 \\
        \hline
        \textbf{Ours} & mel-spectrogram &\textbf{97.2}   \\
        \hline
    \end{tabular}
\end{table}

\begin{figure*}[htbp]
\centering
\begin{tabular}{c}
\includegraphics[height=4.5cm]{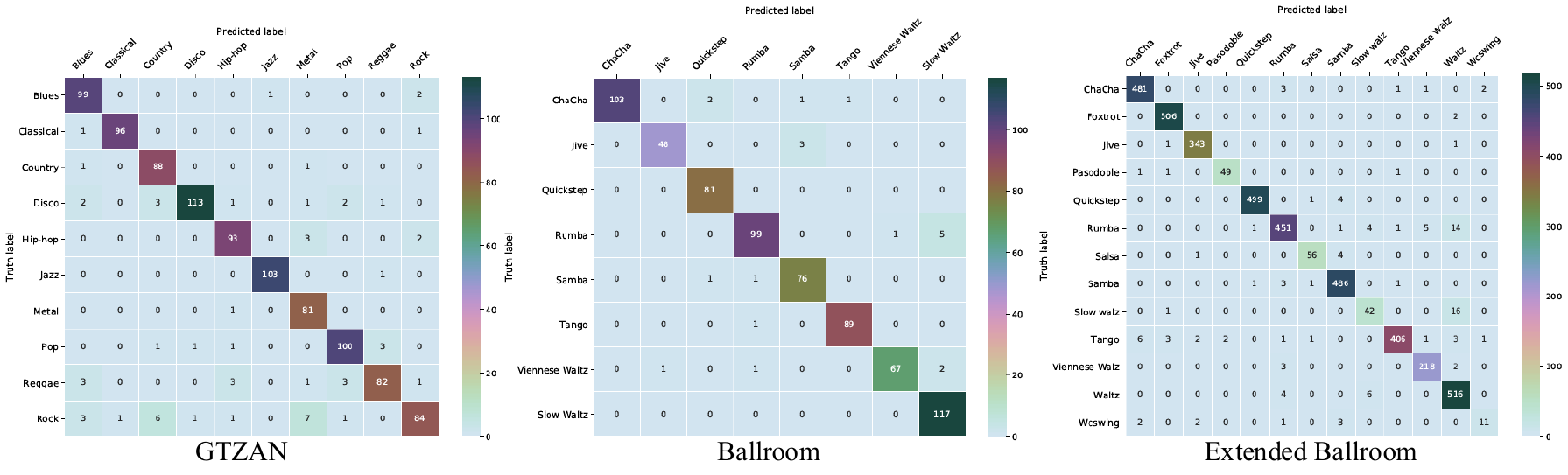}
\end{tabular}
\caption{Accuracies of training, validating, and Testing of each fold on the different datasets}
\label{fig: confusions}
\end{figure*}

Figure~\ref{fig: confusions}(right) presents the confusion matrix of the 10-folds results produced by BBNN model corresponding the accuracy in Table \ref{tab:extended ballroom}. Rows present the dataset ground-truths; Columns denote labels predicted by BBNN model. By analyzing the confusion matrix, it can be noticed that the severe occurrence of confusion is from Rumba and Slow Waltz with Waltz. These genres are difficult to distinguish since they contain similar patterns \citep{lykartsis2015beat}.


Table~\ref{tab:precisions and recall on Extend} reports the tested precision and recall rate of BBNN model for each genre. Since Slow Waltz is prone to misclassified as Waltz, it has a relatively lower precision and recall rate. 
Because the training samples of genre Wcswing are very few, only accounting for 0.5\% of the total sample, the BBNN can only learn severely limited discriminative information. It makes the model's generalization ability for recognizing genre Wcswing relatively poor resulting in the precision and recall rate of Wcswing are relatively low. The recognition ability of BBNN is still robust on the other unbalanced classes such as Pasodoble, Salsa, and Slow Walz, which have slightly more samples than genre Wcswing.
On the whole, the recognition precisions of most genres are more than 90\%, of which Quickstep and Tango are as high as 99\%. 
Based on the results, BBNN has better classification performance even in the case of the incredibly unbalanced dataset.
\begin{table}[htbp]
\centering
\tiny
\caption{Precision (\%) and recall rate (\%) of each genre obtained on the Extend Ballroom dataset}
\label{tab:precisions and recall on Extend}
\begin{tabular}{cccc}
    \hline
    Genre & Precision (\%) & Recall Rate (\%) & F-score (\%) \\
    \hline
    Cha Cha & 98.1 & 98.5 & 98.3 \\
    Foxtrot & 98.8 & 99.6 & 99.2\\
    Jive & 98.5 & 99.4 & 98.9\\
    Pasodoble & 96.0 & 94.2 & 95.1 \\
    Quickstep & 99.6 & 99.0 & 99.3 \\
    Rumba & 96.7 & 94.5 & 95.6 \\
    Salsa & 94.9 & 91.8 & 93.3\\
    Samba & 97.5 & 98.7 & 98.1\\
    Slow Walz &  80.7 & 71.1 &75.6\\
    Tango & 99.0 & 95.3 & 97.1\\
    Viennese Walz & 96.8 & 97.7 & 97.3\\
    Walz & 93.1 & 98.1 &95.5\\
    Wcswing & 78.5 & 57.8 & 66.6 \\
    \hline
    \textbf{Average} & \textbf{94.5} & \textbf{92.0} & \textbf{93.1} \\
    \hline 
\end{tabular}
\end{table}

\section{Conclusions and future work}
\label{sec:conclusions and future work}
In this article, we present a specially designed network for accurately recognizing the music genre. 
The proposed model aims to take full advantage of low-level information of mel-spectrogram for making the classification decision.
We have shown how our model is effective by comparing the state-of-the-art methods, including hand-crafted feature approaches and deep learning models with different architectures, trained on different benchmark datasets. Deep learning approaches usually rely on a large amount of data to train model. In practice, since the number of annotated music recordings per genre class is often limited \citep{fu2011survey}, therefore, except for accuracy, another major challenge is to train a robust CNN model from few labelled data. In this work, the three common datasets (GTZAN, Ballroom, and Extended Ballroom) are employed to validate the proposed network structure from different data scale, especially the Ballroom dataset with only 698 tracks. Experiment results demonstrate that BBNN can overcome this challenging and achieve satisfactory accuracies.

In the future works, we will further improve the proposed model through the following ways. Firstly, we will explore the acoustic features (e.g. SSD, RH) adopting fusion strategies~\citep{wang2017unsupervised, wang2015robust,Wang2016Iterative} as also input into the network. Secondly, in the decision-making stage, we will adopt a new distance metric method~\citep{wang2018multiview} to compute the similarity between genres. 

\bibliographystyle{model2-names}
\bibliography{refs}

\end{document}